\documentclass[10pt]{article}
\usepackage[utf8]{inputenc}
\usepackage[english]{babel}
\usepackage{graphicx}
\usepackage[singlespacing]{setspace}
\usepackage[margin=1in]{geometry}
\graphicspath{{images/}}

\title{Network Analysis of the 2016 Presidential Campaign
  Tweets\footnote{Presented at $4^\mathrm{th}$ International
    Conference on Computational Social Science IC$^{2}$S$^{2}$, July
    12-15, 2018; Northwestern University’s Kellogg School of
    Management, Evanston, IL, USA}}

\author{Dmitry Zinoviev\\\small{Suffolk University, Boston, MA,
    dzinoviev@suffolk.edu}}
 
\date{}

\begin{document}

\maketitle

\begin{abstract}
  We applied complex network analysis to $\sim27,000$ tweets posted by
  the 2016 presidential election's principal participants in the
  USA. We identified the stages of the election campaigns and the
  recurring topics addressed by the candidates. Finally, we revealed
  the leader-follower relationships between the candidates. We
  conclude that Secretary Hillary Clinton's Twitter performance was
  subordinate to that of Donald Trump, which may have been one factor
  that led to her electoral defeat.
  \vskip\baselineskip%
\end{abstract}

\section*{Introduction}
Donald Trump has been an avid user of Twitter before, throughout, and
in the aftermath of the 2017 USA presidential election
campaign. Secretary Hillary Clinton, the Democratic Party candidate,
was active on Twitter only from the beginning to the end of the
campaign. This paper aims to reconstruct the campaign's timeline and
logic using complex network analysis of the tweets posted by Hillary
Clinton, Ted Cruz, John Kasich, Marco Rubio, Bernie Sanders, Jill
Stein, and Donald Trump.

\section*{Dataset}
\begin{table}[t]\centering
  \caption{\label{table}Dataset Summary. $^{**}$More data used in the pilot study. $^*$The limit of the observation period.}\vskip0.5\baselineskip
  \begin{tabular}{lrrlcl}\hline    
    Candidate& \# of Tweets & Words per Tweet&&Campaign&\\
    &&& Start & Length (months) & End\\\hline
    Trump$^{**}$ & 9,984&22.4&Mar 2015$^*$&22& Dec 2016$^*$\\
    Stein & 6,215 & 26.6 & Jul 2015 & 12 & Dec 2016$^*$\\
    Clinton$^{**}$ & 6,082 & 21.6 & Apr 2015 & 20 & Nov 2016\\
    Cruz & 4,501 & 17.6 & Mar 2015$^*$ & 21 &Nov 2016 \\
    Sanders & 4,385& 25.4 & Mar 2015$^*$ & 22 &Dec 2016$^*$\\
    Kasich & 3,687 & 20.8 & Mar 2015$^*$ & 21 &Dec 2016$^*$\\
    Rubio & 2,933 & 19.0 & Apr 2015 & 15 & Aug 2016\\\hline
    Total: & 26,733&&&&
    \\\hline
\end{tabular}
\end{table}

We analyzed $\sim$27,000 tweets posted by Donald Trump and his
contenders in January 2014--December 2017, covering the three major
stages of their quest for the Presidency (Table~\ref{table}). The
tweets are composed of 14,373 distinct, unstemmed terms (words, word
combinations, and hashtags). We selected only 300 base terms that
occurred in the whole corpus at least 100 times for further
analysis. We grouped the tweets by months into 48 tweet corpora and
converted them into a network, based on the vocabulary's
similarity. Each network node represents a monthly corpus. We
connected two nodes with a weighted edge if the base terms'
frequencies in the corresponding corpora were strongly correlated
(with the Pearson correlation coefficient, serving as the edge's
weight, at or above 0.6).

\section*{Network of Tweets}
\begin{figure}[b!]
\centering
\includegraphics[width=\textwidth]{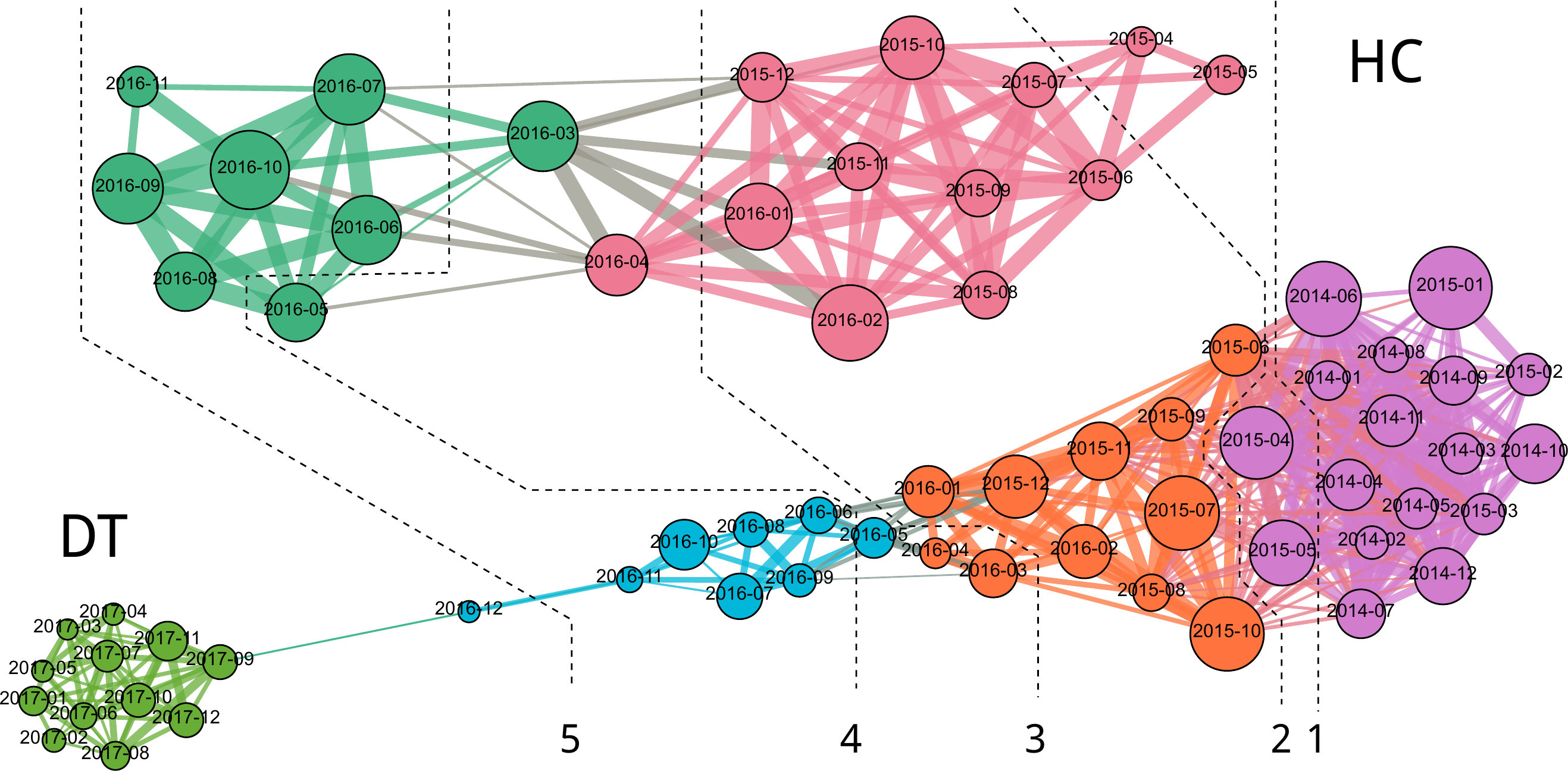}
\caption{Twitter-based electoral campain timelime of Donald Trump (DT)
  and Hilary Clinton (HC) and its stages: 1.~Clinton intends to
  run. 2.~Trump intends to run. 3.~Republican primaries
  start. 4.~Trump's nomination.  5.~General election}
\label{fig:image}
\end{figure}

The resulting network of monthly tweet corpora
(Figure~\ref{fig:image}) has an excellent community structure (with
the Newman modularity $m=0.47$~\cite{newman2006}) that we revealed by
applying the Louvain community detection
algorithm~\cite{blondel08}. We identified four distinct node
clusters. The nodes in each cluster form a temporal continuum and
correspond to the four campaign stages: January 2014--June 2015 (when
Trump announced his intention to run); July 2015--April 2016; May 2016
(when Trump de facto won the Republican Party primary) to November
2016 (the general election); and December 2016--December 2017. We also
observed that the average number of tweets per corpus decreased over
time: from ca.~2,000 for the first two clusters to ca.~1,325 and
ca.~870 for the next two clusters (the differences between the second
and third and third and fourth clusters are statistically
significant).

The resulting network is quasi-linear: 75\% of the nodes are most
strongly connected to the nodes representing the next or previous
month, suggesting that, in general, the campaign timeline was
semantically smooth. Only five nodes' closest neighbors are more than
two months apart.

In particular, we noticed the lack of correlation between December
2016 and January 2017 (when Trump officially became the
POTUS). Instead, the December 2016 corpus correlates well with the
corpora representing the second half of 2017 (July through
October). We hypothesize that the discontinuity was caused by the
unanticipated resistance from the judicial and legislative branches
that Trump faced immediately after taking office, which put him in the
uphill fight mode.  Only in the second half of the year, especially
after securing the Congress support for the tax reform, Trump
realigned his tweets' rhetorics. We will further look into more
fine-grain, bi-weekly 2017 tweet corpora and their vocabulary to
investigate this hypothesis.

We observed that the well-structured, quasi-linear network of monthly
Trump tweet corpora matches the presidential election campaign's
calendar stages. To provide additional support to this observation, we
applied the same analysis to Clinton's (the other major participant of
the campaign) tweets. Clinton's Twitter corpus consists of ca.~6,400 tweets,
99\% of which were posted between April 2015 and November 2016.

The corpus consists of two clusters: April 2015 (when Clinton
announced her intention to run)--March 2016 and April 2016--November
2016 (the general election). Interestingly, the boundary between the
clusters corresponds to the campaign stages, too---but those of Trump
rather than Clinton. It is located in the middle of the Republican
Party's primary elections. 

\section*{Topics Extraction}
We hypothesize that Hillary Clinton was in the defensive position
throughout the campaign and had to respond to Trump's challenges,
rather than form her agenda (at least on Twitter). To confirm our
hypothesis, we look into the cross-correlations between the major
presidential candidates' tweet corpora to better understand their
``leader-follower'' relationship.
\begin{figure}[tb]
\centering
\includegraphics[width=\textwidth]{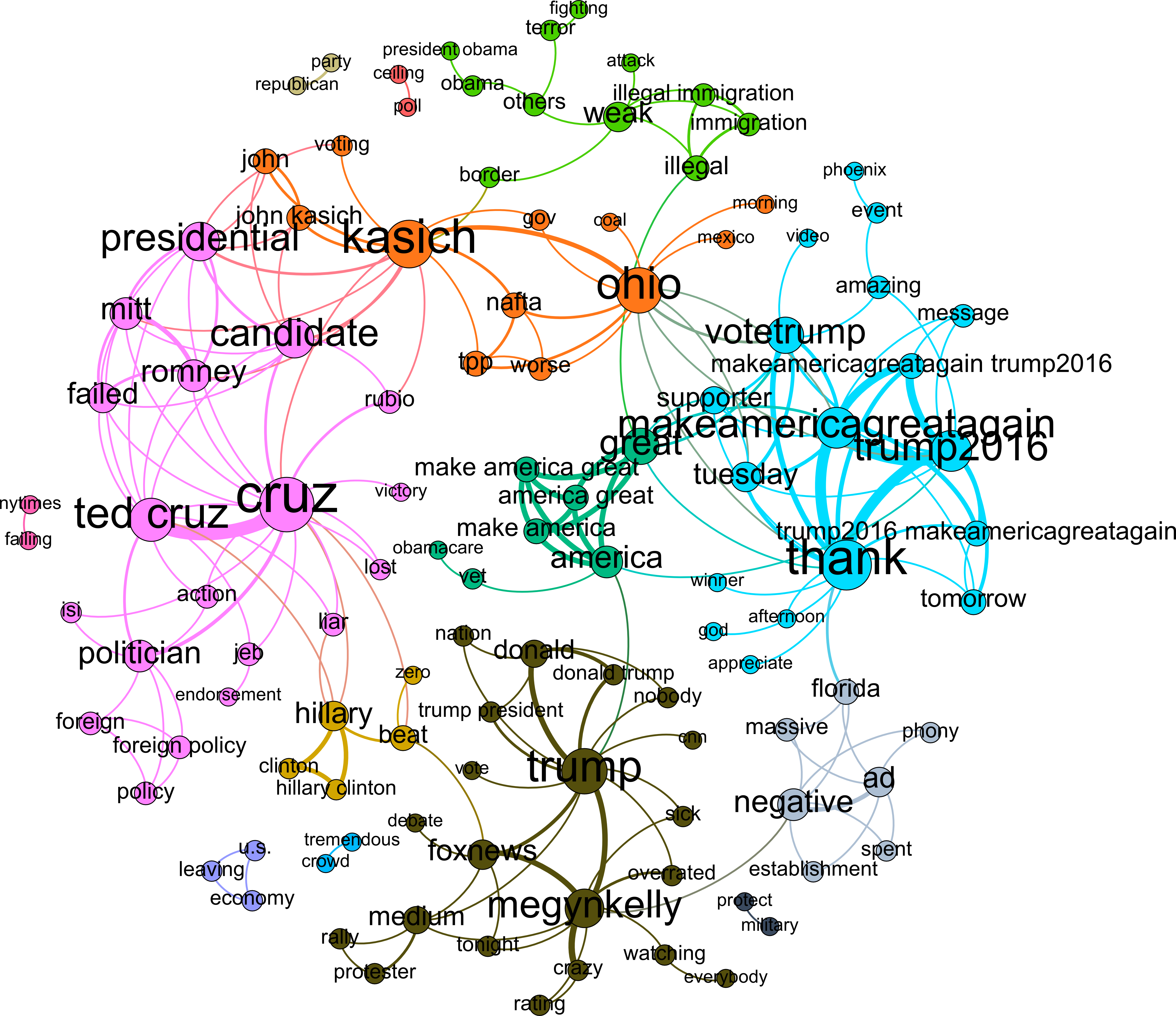}
\caption{Sample semantic network for Donald Trump, the second half of
  March 2016.}
\label{fig:sample_network}
\end{figure}

We automatically extracted the terms (hashtags, rare and significant
words, and their bigrams and trigrams) from the tweets. We defined
word eligibility based on the Corpus of Contemporary American English
(COCA): a word is rare if it is not in the COCA or its frequency is
less than 25$\times mean(COCA)$. A word is significant if it occurs at
least ten times, and its frequency is greater than 25$\times
mean(tweets)$. We identified 1,416 terms, such as ``-john,''
``2-party,'' ``2016,'' ``47246,'' ``ability,'' ``abolish,''
``abolishtheirs,'', ``abortion,'' ``absolutely,'' ``absurd,'' \ldots,
``worried,'' ``worse,'' ``worst,'' ``worth,'' ``written,'' ``y'all,''
``yesterday,'' ``youth,'' ``zero,'' and ``---hillary.''

For each candidate $C$ and each biweekly subcorpus $W$, we created 287
semantic networks of terms by treating the terms as nodes and their
co-occurrence in the same tweet as edge weight. We applied the Louvain
algorithm~\cite{blondel08} to identify communities of terms addressed
by each candidate over each two-week span, and ended up with 2,518
subtopics $ST\left(C,W,i\right)$, where $i$ is the subtopic
identifier. Figure~\ref{fig:sample_network} shows the sample semantic
network for Donald Trump for the second half of March 2016. Note how
he chiefly mentions Cruz and Kasich, merely notices Clinton, and
ignores the other candidates.

We combined the subtopics that have a significant word overlap by
constructing a weighted network of subtopics. We treat each subtopic
as a node and use the Jaccard index as edge weight. We again apply the
Louvain algorithm and identify seventeen recurring subtopics for all
candidates and periods---the topics ``as such.'' Since a term may
belong to more than one topic, we treat a topic as a fuzzy set. The
following list shows the partial compositions of each topic in the
order of decreasing frequency.

\begin{enumerate}
\item hillary; campaign; trump; clinton; vote\ldots
\item america; great; make america; america great; make america
  great\ldots
\item obama; iran; immigration; president obama; terrorism\ldots
\item tonight; tune; sanders; town; town hall\ldots
\item wage; minimum; minimum wage; worker; living\ldots
\item forward; looking; looking forward; tomorrow; ticket\ldots
\item tax; policy; foreign; foreign policy; military\ldots
\item health; health care; obamacare; republican; drug\ldots
\item student; debt; student debt; economy; wall street\ldots
\item climate; climate change; fossil fuel; fuel; energy\ldots
\item failing; prayer; nytimes; thought; dishonest\ldots
\item justice; criminal; police; equality; officer\ldots
\item united; united states; citizen; america
\item happy; birthday
\item conservative; courageous conservative; courageous
\item lesser evil; evil; lesser; greater
\item god; bless.
\end{enumerate}

\section*{Leadership and Followership}

\begin{table}[tb!]\centering
  \caption{\label{table:topic10}Leaders and Followers in the Topic \#10
    ``Climate Change.''}\vskip0.5\baselineskip
  \begin{tabular}{lll}\hline
    Timespan & Partricipants/Leaders & Newcomers/Followers \\\hline
    2015-08-02 & Sanders, Trump & Clinton\\
    2015-09-13 & Sanders & Rubio\\
    2015-09-27 & Sanders, Rubio & Clinton\\
    2015-10-11 & Clinton & Sanders\\
    $\ldots$ & $\ldots$ &$\ldots$ \\
    2016-07-03 & Sanders, Stein & Clinton\\
    2016-09-11 &  Sanders, Stein & Rubio\\    
    2016-10-23 & Stein & Sanders, Clinton \\
    2017-01-01 & Sanders & Stein
    \\\hline
  \end{tabular}
\end{table}

To track topics' popularity, we looked at each recurring topic, each
candidate, and each period, and calculated the Pearson correlation
$\rho$ between the frequencies of the terms in the topic and in the
subtopic for that span. If the correlation is high ($\rho\gg0$), then
the candidate consistently used the topic vocabulary over the time
span. If the correlation is low ($\rho\approx0$), then the candidate
disregarded the topic vocabulary over the period.
To track followers and leaders, we check, for each candidate, topic,
and period, whether the candidate covered the topic during that
period, but not during the previous period (two weeks ago). If so,
then the candidate followed other candidates who had covered the same
topic two weeks ago. Those candidates are the leaders. (See
Table~\ref{table:topic10}.)

\begin{figure}[b!]
\centering
\includegraphics[width=.65\textwidth]{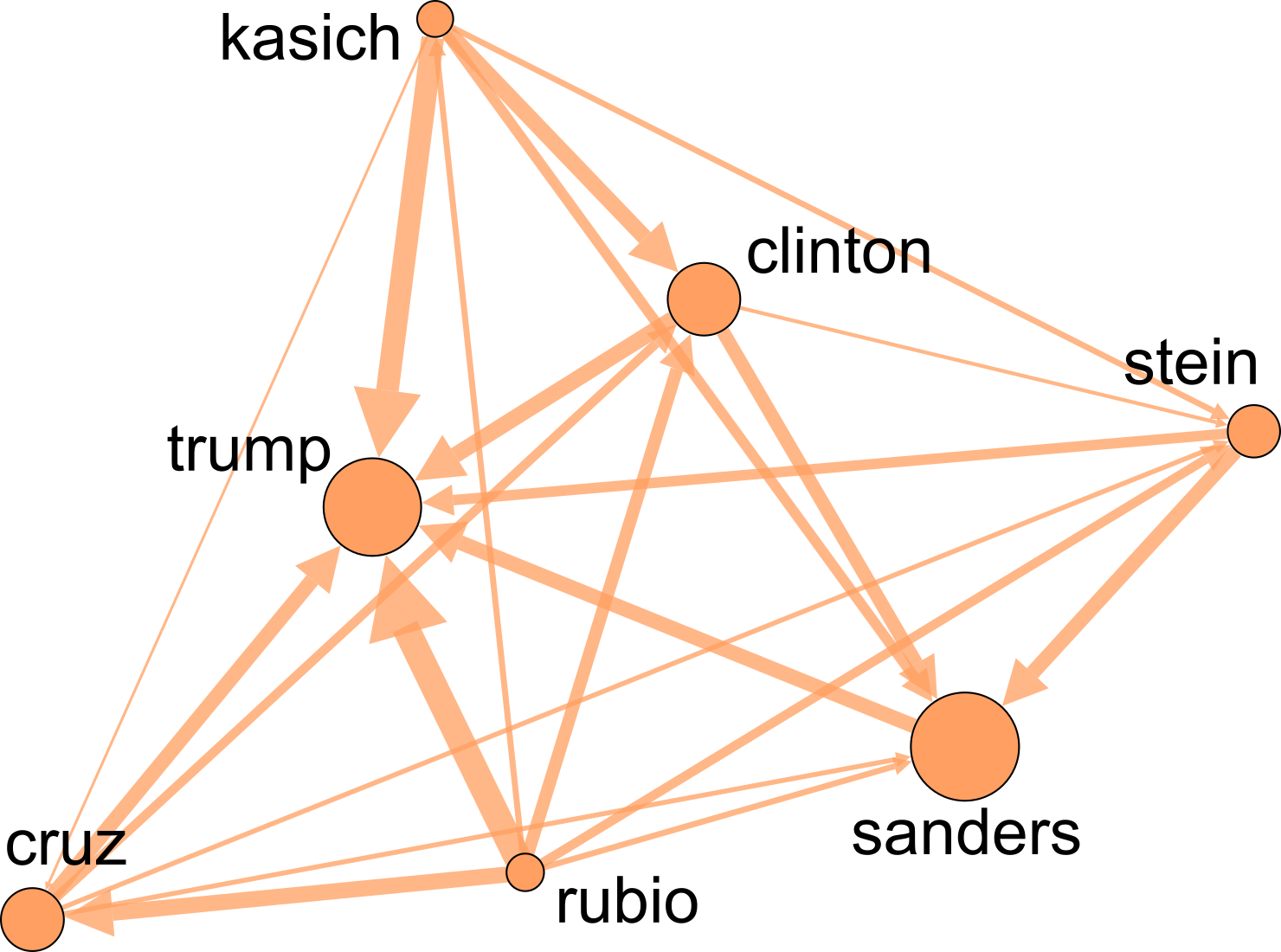}
\caption{The network of leaders and followers.}
\label{fig:followers}
\end{figure}

As a result, we can construct yet another weighted directed network
bases on who followed whom and how many times
(Figure~\ref{fig:followers}). The network clearly illustrates that the
most followed presidential candidate was Donald Trump. He rarely
followed anyone else, unlike his principal competitors: Hillary
Clinton and Bernie Sanders. The least followed candidates are John
Kasich and Marco Rubio.

Finally, we introduce measures of engagement and leadership. We define
leadership as the number of biweekly spans when a candidate is
followed by other candidates, minus the number of spans when the
candidate follows other candidates. Leadership can be positive or
negative. Similarly, we define engagements and the total number of
spans when the candidate follows or is being followed---that is, is
engaged in an implicit or explicit conversation with the other
candidates. The engagement is always strictly positive.

\begin{figure}[tb]
\centering
\includegraphics[width=.7\textwidth]{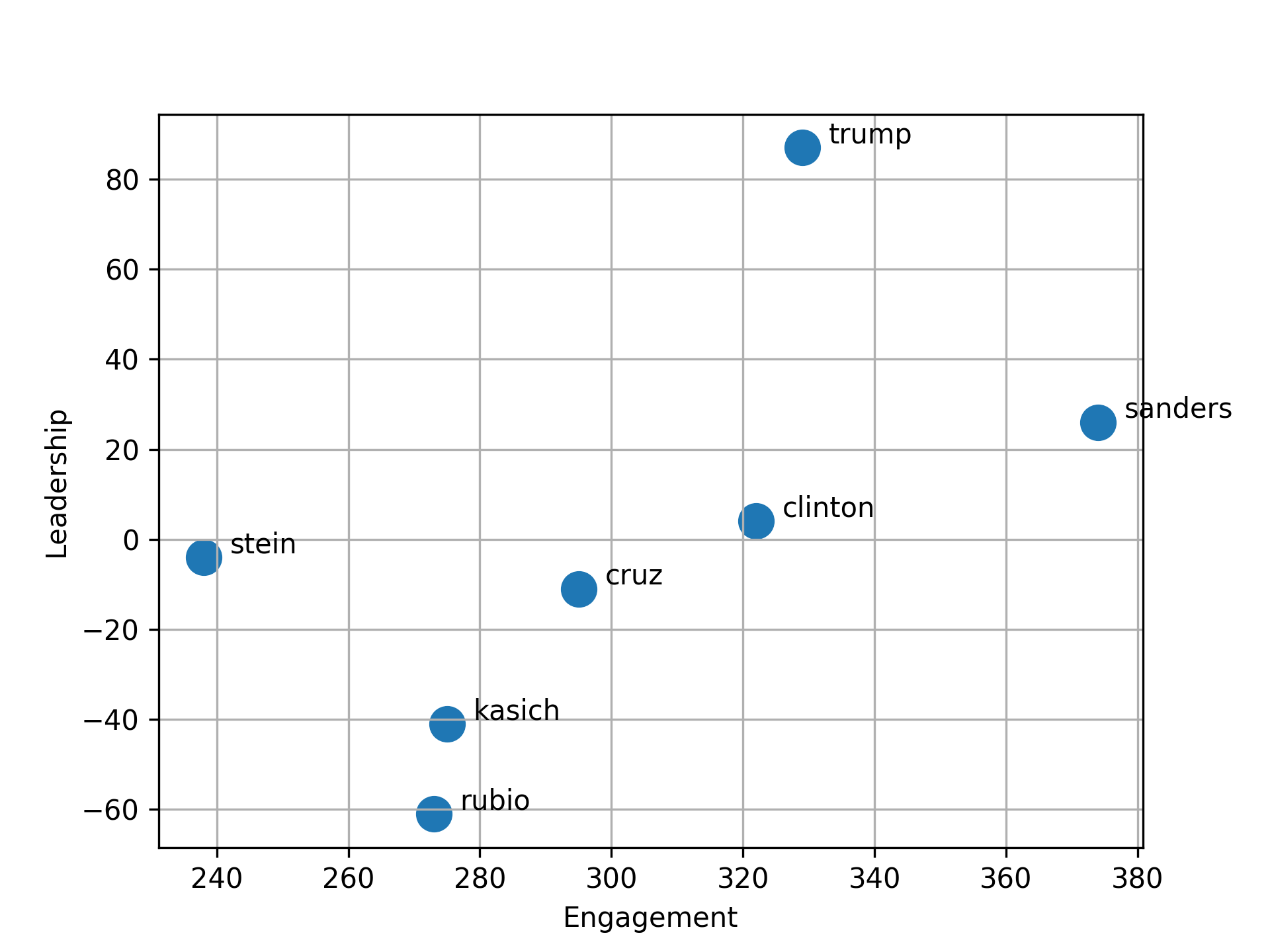}
\caption{Engagement and leadership scores of the presidential candidates.}
\label{fig:followers+leaders}
\end{figure}

Figure~\ref{fig:followers+leaders} shows the leadership and engagement
scores for each candidate. Three candidates stand apart from the pack:
Donald Trump with the highest leadership score (consistent with
Figure~\ref{fig:followers}), Bernie Sanders with the highest level of
engagement, and Jill Stein with a shallow level of engagement.

\section*{Conclusion}

We applied complex network analysis to the tweets posted by the
principal participants of the 2016 presidential election campaign in
the USA. We revealed recurrent topics of the candidates' tweets and
the leader/follower relationships between the candidates. The number
of the recurring topics owned or shared by the candidates over the
campaign timeline is low: only seventeen, dealing with President
Obama's heritage, America in general, healthcare, minimum wages,
taxation, student debt, etc. Trump and Sanders were the most followed
candidates, while Kasich and Rubio were the least followed
candidates. Simultaneously, Trump was the most followed and the second
most engaged candidate, which may have partially redetermined his
victory.

\section*{Acknowledgment}
The author is grateful to Elena Llaudet (Suffolk University, Boston)
for her helpful suggestions.

\bibliographystyle{acm}
\bibliography{cs}

\begin{thebibliography}{1}

\bibitem{blondel08}
{\sc Blondel, V., Guillaume, J.-L., Lambiotte, R., and Lefebvre, E.}
\newblock {Fast Unfolding of Communities in Large Networks}.
\newblock {\em J. of Statistical Mechanics: Theory and Experiment}, 10 (2008),
  1000.

\bibitem{newman2006}
{\sc Newman, M.}
\newblock Modularity and community structure in networks.
\newblock {\em Proceedings of the National Academy of Sciences of the United
  States of America 103}, 23 (2006), 8577--8696.

\end{thebibliography}

\end{document}